\documentstyle[11pt]{article}

\def\beq{\begin{equation}}
\def\eeq{\end{equation}}
\def\pd{\phi_A}
\def\noi{\noindent}
\def\rz{\rho}
\def\rzp{\rho'}
\begin{document}
\begin{center}
{\Large\bf Conformal invariant pomeron interaction in the perturbative
QCD with large $N_c$ }
\vspace{0.5 cm}

M.A.Braun

{\small\it  Department of High
Energy Physics,
University of St. Petersburg,\\ 198904 St. Petersburg, Russia}\\
\end{center}
\vspace{1 cm}

\noi {\bf Abstract}

An effective non-local quantum field theory is constructed, which describes
interaction of pomerons in the high-coloured QCD. The theory includes both
splitting and merging triple pomeron vertexes and diagrams with pomeronic
loops. The Schwinger-Dyson equations for the 'physical' pomeron are
written. Conformal invariance allows to reduce the theory to the old-
fashioned Gribov pomeron theory with an infinite number of pomerons,
one of which is supercritical.

\section{Introduction}
The high-energy behaviour in QCD with a  large number of colours $N_c$
is described by the exchange of hard pomerons, which split and merge
by a triple pomeron vertex. Exchange of coloured objects (single gluons)
is damped by factor $1/N_c^2$. In interaction with heavy nuclei
the leading  contribution comes from diagrams without pomeronic loops
(tree diagrams). Summation of these is achieved by a closed equation for
DIS (the BK equation ~\cite{bal,kov,bra1}) or a closed pair
of equations for nucleus-nucleus scattering ~\cite{bra2}.
Some estimates of the contribution from loops were made in 
~\cite{salam,BRV}.
To take into account the contribution from pomeronic loops
in a consistent manner one has to consider the effective pomeron field 
theory introduced in ~\cite{bra2} as a full-fledged quantum theory. The 
present study is devoted to this aim.

Note that pomeronic loops have lately been also actively
studied in the framework of the colour dipole picture in the
so-called JIMWLK approach
(see e.g.~\cite{jimwlk} and references therin).
There evolution in rapidity of a state considered as a
functional of the gluon field is governed by a certain Hamiltonian
made of the field and functional derivatives in the field.
Taking loops into account
leads to a Hamiltonian containing functional derivatives up to the
fourth order  ~\cite{mul}.
In our approach
a quantum Hamiltonian can also be introduced. However
it contains functional derivatives only up to the second order
and so is considerably simpler than in the JIMWLK approach.

In this study we restricted ourselves to a rather formal treatment
of the perturbative diagrams for the pomeron interaction with or without
loops. We construct the relevant Schwinger-Dyson equations for the
full pomeron Green function and also discuss their conformal 
(Moebius) invariance,
which hopefully may simplify their analysis. In fact we are not very
optimistic about a realistic calculation of the amplitudes
with pomeronic loops included. This problem presented enormous difficulties
already for the much simpler old-fashioned Gribov local supercritical
pomeron model, so that even its internal inconsistency
was claimed ~\cite{abarb}
(see also ~\cite{white} for a discussion of this inconsistency). Possibly 
the more complicated
structure of the BFKL pomeron and its interaction may overcome these
old troubles. However in the present study this problem is not touched
but left for future investigations.
\section{Effective action and diagrams}
\subsection{Effective action}
Our main tool will be the non-forward BFKL Green function
as a function of gluon coordinates and rapidities, which satisfies the
equation
\beq
\Big(\frac{\partial}{\partial}+H\Big)
g(y-y';r_1,r_2;r'_1,r'_2)=
\delta(y-y')\nabla_1^{-2}\nabla_2^{-2}\delta^2(r_{11'})
\delta^2(r_{22'}),
\eeq
where $r_{11'}=r_1-r'_1$ etc and $H$ is the BFKL Hamiltonian
~\cite{lip0}:
\beq
H=\frac{\bar{\alpha}}{2}\Big(\ln p_1^2+\ln p_2^2 +
\frac{1}{p_1^2}\ln r_{12}^2\cdot p_1^2+
\frac{1}{p_2^2}\ln r_{12}^2\cdot p_2^2-4\psi(1)\Big)
\eeq
and $\bar{\alpha}=\alpha_sN_c/\pi$.
To economize on notations,
in the following we  shall denote
as $z$ the point in the space formed by rapidity $y$ and two transverse
vectors $r_1,r_2$:
\[ z=\{y, r_1,r_2\}=\{y,\rz\}.\]
In this notation we write $g(y-y';r_1,r_2;r'_1,r'_2)=g(z,z')$. The Green
function
$g(z,z')$ is invariant under conformal (Moebius) transformations of
coordinates..

The inverse function $g^{-1}(z,z')$ defined by
\beq
\int dz''g^{-1}(z,z'')g(z'',z')=\delta(z-z'),
\eeq
where $dz=dyd^2r_1d^2r_2\equiv dyd\rho$ and
$\delta(z)=\delta(y)\delta^2(r_1)\delta^2(r_2)\equiv\delta(y)\delta\rho$,
is however not conformal invariant, since
the measure is not conformal invariant.
It is not difficult to construct a conformal invariant inverse $g_{inv}^{-1}$
using the invariant measure:
\beq
d\tau=\frac{d^2r_1d^2r_2}{r_{12}^4}.
\eeq
Then we can rewrite (3) as
\beq
\int dy''d\tau''
r_{12}^4g^{-1}(z,z''){r''_{12}}^4g(z'',z')= r_{12}^4\delta(z-z').
\eeq
This shows that the conformal invariant function is
\beq
g_{inv}^{-1}(z,z')=r_{12}^4g^{-1}(z,z'){r'_{12}}^4
\eeq

Now we pass to constructing the effective  non-local
quantum field theory, which is to describe propagation of free
pomerons and their triple interaction. It has  to
generate all diagrams built from the pomeron propagator, which is
the Green function $g$, and triple pomeron vertex introduced in
~\cite{BW,AHM,BV}.  Let $\phi_B(z)$ and $\pd(z)$ be
two bilocal fields for the incoming and outgoing pomerons.
To reproduce the correct propagator the free action $S_0$ has to be
\beq
S_0=\int dzdz'\pd (z)g^{-1}(z,z')\phi_B(z).
\eeq
We introduce interaction with an external sources as
\beq
S_E=-\int dz \pd (z)J_B(z)+\Big(A\leftrightarrow B\Big).
\eeq
Here the sources for the projectile A and target B are different from
zero at rapidities
$y=Y$ and $y=0$ respectively:
\beq
J_A(z)=\bar{J}_A(\rho)\delta(y-Y),\ \ J_B(z)=\bar{J}_B(\rho)\delta(y).
\eeq
Finally the interaction  to reproduce the 3P vertex at large $N_c$ has to 
be taken in the
form:
\beq
S_I=\frac{2\alpha_s^2N_c}{\pi}
\int dy\frac{d^2r_1d^2r_2d^2r_3}{r_{12}^2r_{23}^2r_{31}^2}
\phi_B(z_1)\phi_B(z_2) L_{12}\pd (z_3) +\Big(A\leftrightarrow B\Big),
\eeq
where $z_1=\{y,r_2,r_3\}$,  $z_2=\{y,r_3,r_1\}$, $z_3=\{y,r_1,r_2\}$
and the conformal invariant operator $L_{12}$ is
\beq
L_{12}=r_{12}^4\nabla_1^2\nabla_2^2.
\eeq
Note that the form (10) assumes the fields to be symmetric in the two
space points $r_1$ and $r_2$. In fact the symmetry properties of the
fields
are determined by symmetry properties of the external sources. We assume them
to be symmetric under the interchange $r_1\leftrightarrow r_2$.

This action leads to BFKL pomeron diagrams with the standard 3P
interaction in the presence of an external field. Note that the signs of
different building blocks of the diagrams is somewhat different
from the standard ones: the propagator $g$ and external sources enter
with a minus sign as a consequence of the choice of signs in the action.
This latter corresponds to the desire to make the interaction real
and not pure imaginary as in the original Gribov reggeon field theory.

In absence of the external sources this action is explicitly
conformal invariant provided $\phi_B$ and $\pd$ are  invariant.
Indeed the free part can be rewritten as
 \beq S_0=\int
dydy'd\tau d\tau'\pd
(z)r_{12}^4g^{-1}(z,z'){{r'_{12}}^4}\phi_B(z)
\eeq
and the
interaction part as
\beq
 S_I=\frac{2\alpha_s^2N_c}{\pi} \int
dydy'dy''d\tau d\tau' d\tau''
\phi_B(z')\phi_b(z'')\gamma(z',z''|z)L_{12}\pd (z) +
\Big( A\leftrightarrow  B\Big),
\eeq
 where $\gamma (z',z''|z)$ is the
bare interaction vertex for the incoming pomeron at $z$ and two
outgoing pomerons at $z'$ and $z''$:
\beq
\gamma(z',z''|z)=\delta(y-y')\delta(y-y'')
\delta^2(r_{12'})\delta^2(r_{1'2''})\delta^2(r_{1''2})
r_{12}^2r_{1'2'}^2r_{1''2''}^2.
\eeq
This is a conformal invariant
function. Since both $r_{12}^4g^{-1}(z,z'){{r'_{12}}^4}$ and
$\gamma(z',z''|z)$ are conformal invariant, so is the action
$S_0+S_I$. Of course in  physically relevant cases  conformal
invariance is always broken by the external sources, which also
introduce a mass scale into the theory. As a result contribution
from any Feynman diagram without external sources is conformal
invariant, which becomes explicit if one uses the invariant
integration measure $dz_{inv}=dyd\tau$ and invariant
interaction vertex $\gamma$ given by Eq. (14).

Finally note that from (1) it follows that as an operator in the $z$-space
\beq
g^{-1}(z,z')=\nabla_{\rho}^2\left(\frac{\partial}{\partial y}+H\right).
\eeq
Here we use the notation
\beq
\nabla_{\rho}^2=\nabla_1^2\nabla_2^2.
\eeq
Note that operator $\nabla_{\rho}^2\,H$ is symmetric:
\beq
\nabla_{\rho}^2\, H=
\frac{\bar{\alpha}}{2}\Big(p_1^2p_2^2\ln (p_1^2 p_2^2) +
p_2^2\ln r_{12}^2p_1^2+
p_1^2\ln r_{12}^2p_2^2-4\psi(1)p_1^2p_2^2\Big).
\eeq
As a result $S_0$ can be written directly in terms of the BFKL
Hamiltonian:
\beq
S_0=\int dy d\rho \phi_A(y,\rho)\nabla_{\rho}^2
\left(\frac{\partial}{\partial y}+H\right)\phi_B(y,\rho)=
\int dy d\rho \phi_B(y,\rho)\nabla_{\rho}^2
\left(-\frac{\partial}{\partial y}+H\right)\phi_A(y,\rho).
\eeq
In this form  the symmetry between the projectile and target is
made explicit: it has to be accompanied by changing
$y\to -y$.

Using a representation for $H$ (~\cite{BLV})
\beq
Hf(r_1,r_2)=\frac{\bar{\alpha}}{2\pi}\int \frac{d^2r_3
r_{12}^2}{r_{13}^2r_{23}^2}\Big(f(r_1,r_2)-f(r_1,r_3)-f(r_2,r_3)\Big)
\eeq
we can rewrite the free part of the action in a more explicit form
\[
S_0=\int dy
d^2r_1d^2r_2\pd(y,r_1,r_2)\nabla_{\rho}^2\Big\{\frac{\partial}
{\partial y}\phi_B(y,r_1,r_2)\]\beq +\frac{\bar{\alpha}}{2\pi}\int \frac{d^2r_3
r_{12}^2}{r_{13}^2r_{23}^2}\Big(\phi_B(y,r_1,r_2)-
\phi_B(y,r_1,r_3)-\phi_B(y,r_2,r_3)\Big)
\Big\}.
\eeq

\subsection{Diagrams and their order of magnitude}
The quantum field theory described by the action $S=S_0+S_I+S_E$
allows to construct the perturbation theory by the standard technique,
expressing the amplitude as a sum of Feynman diagrams. It is instructive to
see orders of magnitude of different contributions in terms of two
independent small parameters of the theory, $ \bar{\alpha}=\alpha_sN_c/\pi$
and $1/N_c$. Obviously each triple interaction contributes
$\bar{\alpha}^2/N_c$. The final estimate depends on the magnitude of the
external source. If we treat it perturbatively to be consistent with the
whole approach then it should correspond to the quark-antiquark loop,
from which we extract its order, which is $\bar{\alpha}$.
As a result, the diagram with $l_E$ external lines (sources) and $L$
loops has the order
\beq
\bar{\alpha}^{l_E}\Big(\frac{\bar{\alpha}^2}{N_c}\Big)^{2L+l_E-2}.
\eeq
As one sees, for a given number of external sources, the dominant
contribution comes from the tree diagrams, each loop introducing
a small factor $\bar{\alpha}^4/N_c^2$.

However this pure counting does not take into account the growth of
the pomeron propagator at large $y$ as $\exp \Delta y$ where
$\Delta$ is the BFKL intercept, nor the enhancement related to
the nuclear sources with large atomic numbers. For
the AA collision amplitude with the overall
rapidity difference $Y$ this changes (21) to
\beq
\Big(\bar{\alpha} A^{1/3}\Big)^{l_E}
\Big(\frac{\bar{\alpha}^2}{N_c}\Big)^{2L+l_E-2}e^{n\Delta Y}.
\eeq
where $n$ is the maximal number of the exchanged pomerons at a given
rapidity. This number depends on the topology of the diagram and
generally grows with $l_E$ and $L$.

So for
\beq
\bar{\alpha} A^{1/3} \sim 1
\eeq
one has to sum all tree diagrams and
for rapidities $Y$ such that
\beq
\frac{\bar{\alpha}^2}{N_c}e^{\Delta Y}\sim 1
\eeq
one has to sum also at least some of the loops.
For onium-onium scattering ($l_E=2$) a convenient method to sum the
leading contribution is to join two sums of tree diagrams starting
from the projectile and target at mid-rapidity ~\cite{salam}.

However all these perturbative estimates are to be taken with caution.
The full pomeron Green function (sum of the so-called enhanced diagrams)
may have an asymptotic behaviour at large $y$ very different from the
bare one. In fact there is every reason to believe that the former
will grow at most as a power of $y$, not as an exponential.
Then all the above estimates will have to be reconsidered. Unfortunately
at present a reliable summation of all enhanced diagrams cannot be 
performed even for the old Gribov local supercritical pomeron model.

\section{The pomeron Hamiltonian and operators}

With the action fixed, one can easily construct a Hamiltonian
formulation for the state
evolution.
Since the Lagrangian is of the first order in derivatives in rapidity
the Hamiltonian is just  the action without  the derivative terms
with a minus sign and integration over $y$ dropped:
\[
{\cal H}=-
\int d\rho \pd(\rho)\nabla_{\rho}^2H\phi_B(\rho)\]\beq
-\frac{2\alpha_s^2N_c}{\pi}\int\frac{d^2r_1d^2r_2d^2r_3}{r_{12}^2r_{23}^2
r_{31}^2}
\Big\{\phi_B(\rho_{23})\phi_B(\rho_{31})L_{12}\pd(\rho_{12})+
\Big(A\leftrightarrow B\Big)\Big\},
\eeq
where $\rho_{23}=\{r_2,r_3\}$ etc.
To pass to the quantum theory one has to consider $\phi_{A,B}(\rho)$ as
operators.
Their commutation relation can be easily
established from the form of the free Green function
\beq
g(y-y';\rho,\rho')=-<T\{\phi_B(y,\rho)\pd(y',\rho')\}>.
\eeq
From (26) we conclude
\beq
\Big(\frac{\partial}{\partial y}+H\Big)
<T\{\phi_B(y,\rho)\pd(y',\rho')>=\delta(y-y')[\phi_B(y,\rho),
\pd(y,\rho')],
\eeq
where we have used the equation of motion for $\phi_B$.
Comparison with (1) gives
\beq
[\phi_B(y,\rho),\pd(y,\rho')]=-
\nabla_{\rho}^{-2}\delta(\rho-\rho').
\eeq

Thus in a representation in which $\phi_B(\rho)$ is diagonal and the state
vector is a functional
$
\Psi\{\phi_B(\rho)\}
$
the field $\phi_A$ is essentially a functional derivative
\beq
\pd(\rho)=\nabla_{\rho}^{-2}\frac{\delta}{\delta \phi_B(\rho)}.
\eeq
The state with a given field $\pd (\rho)$ will be represented by a
an exponential
\beq
\Psi_{\pd(\rho)}(\{\phi_B\})
=e^{\int d\rho\phi_B(\rho )\nabla_{\rho}^2\pd(\rho)}.
\eeq

The state vector will satisfy the evolution equation
\beq
\frac{d\Psi}{dy}={\cal H}\Psi,
\eeq
where in ${\cal H}_B$, given by (25), one has to substitute
the field $\pd$ by functional derivatives.
In this substitution, as always, the order of the operators
is actually undetermined. If we put all the derivatives to the right
('normal ordering') then explicitly
\[
{\cal H}=
-\frac{\bar{\alpha}}{2\pi}\int
\frac{d^2r_1d^2r_2d^2r_3r_{12}^2}{r_{13}^2r_{23}^2}
\Big\{\Big[
\phi_B(\rho_{12})-\phi_B(\rho_{13})-\phi_B(\rho_{23})\Big]
\frac{\delta}{\delta \phi_B(\rho_{12})}\]\beq-
4\pi\alpha_s\Big[\phi_B(\rho_{13})\phi_B(\rho_{23})
\frac{\delta}{\delta \phi_B(\rho_{12})}+L_{12}\phi_B(\rho_{12})
\Big(\nabla_{\rho}^{-2}\frac{\delta}{\delta \phi_B(\rho_{13})}\Big)
\Big(\nabla_{\rho}^{-2}\frac{\delta}{\delta \phi_B(\rho_{23})}\Big)
\Big]\Big\}.
\eeq

One can pass from this 'target' representation in which the
field $\phi_B$ is diagonal to the 'projectile' representation in which
it is $\phi_A$ which is diagonal and $\phi_B$ is represented by a functional
derivative
\beq
\phi_B(\rho)=-\nabla_{\rho}^{-2}\frac{\delta}{\delta \phi_A(\rho)}
\eeq
In this representation the Hamiltonian will be obtained from (32) by interchanging
$A$ and $B$ and changing signs of derivatives. The state vector will be obtained
by a quasi-Fourier transformation using (30).

One can construct a formulation in which the symmetry between target and projectile
is more explicit. To do this one can pass to slightly different
field variables for which the BFKL Hamiltonian becomes symmetric.
The form of these new variables is clearly seen from the commutation
relation (28)
\[
\phi(y,\rho)=\sqrt{\nabla_{\rho}^2}\phi_A(y,\rho)
\equiv T\phi_A(y,\rho),\]\beq
\phi^{\dagger}(y,\rho)=\sqrt{\nabla_{\rho}^2}\phi_B(y,\rho)
\equiv T\phi_B(y,\rho).
\eeq
For them the equal rapidity commutation relation takes the form
\beq
[\phi(y,\rz),\phi^{\dagger}(y,\rzp)=\delta(\rz-\rzp).
\eeq
One can also assume that the scalar product of state vectors is chosen to make
$\phi$ and $\phi^{\dagger}$ Hermitian conjugate to each other.
In the representation in which say $\phi^{\dagger}$ is diagonal with complex
eigenvalues $\alpha$ we take (up to a normalization factor)
\beq
<\Psi_1|\Psi_2>=\int D\alpha D\alpha^*\, \Psi_1(\alpha^*)\Psi_2(\alpha)
e^{-\int d\rho\alpha^*(\rho)\alpha(\rho)}.
\eeq
Then indeed
\[
<\Psi_1|\phi^{\dagger}(\rho)|\Psi_2>=\int D\alpha D\alpha^*\, \Psi_1(\alpha^*)\Psi_2(\alpha)
\alpha(\rho) e^{-\int d\rho'\alpha^*(\rho')\alpha(\rho')}\]\[=
\int D\alpha D\alpha^*\,\Psi_1(\alpha^*)\Psi_2(\alpha)
\Big(-\frac{\delta}{\delta \alpha^*(\rho)}
\Big)e^{-\int d\rho'\alpha^*(\rho')\alpha(\rho')}
\]\[=\int D\alpha D\alpha^*\, \Psi_2(\alpha)
e^{-\int d\rho'\alpha^*(\rho')\alpha(\rho')}
\frac{\delta}{\delta \alpha^*(\rho)}\Psi_1(\alpha^*)
=<\phi(\rho)\Psi_1|\Psi_2>.\]
Thus the two quantized fields $\phi(y,\rz)$ and $\phi^{\dagger}(y,\rz)$
acquire the standard meaning of annihilation and creation operators for a
pomeron at rapidity $y$ and space points  $\rz=\{r_1,r_2\}$.

In terms of these new field variables the free action takes the form
\beq
S_0=\int dy d\rho\phi(\rho)T
\left(\frac{\partial}{\partial y}+H\right)T^{-1}\phi^{\dagger}(\rho)\equiv
\int dy d\rho\phi(\rho)
\left(\frac{\partial}{\partial y}+\bar{H}\right)\phi^{\dagger}(\rho),
\eeq
where  a new Hamiltonian for the pomeron is
\beq
\bar{H}=THT^{-1}=
\frac{\bar{\alpha}}{2}\Big(\ln p_1^2+\ln p_2^2 +
\sqrt{\frac{p_2^2}{p_1^2}}\ln r_{12}^2\sqrt{\frac{p_1^2}{p_2^2}}+
\sqrt{\frac{p_1^2}{p_2^2}}\ln r_{12}^2\sqrt{\frac{p_2^2}{p_1^2}}
-4\psi(1)\Big).
\eeq
It has obviously the same eigenvalues but is Hermitian
(and real). Using its hermiticity we can revert the order of operators in
(37) and write $S_0$ in the 'normal order' form
\beq
S_0=
\int dy d\rho\phi^{\dagger}(\rho)
\left(-\frac{\partial}{\partial y}+\bar{H}\right)\phi(\rho).
\eeq

This form explicitly shows the symmetry between target and projectile, which
is quite similar to the usual time reversal: one has to change
$\phi\leftrightarrow \phi^{\dagger}$, $y\to -y$ and revert the order of
all operators.

In terms of new field operators the external part of the action aquires the
form
\beq
S_E=-\int dz \Big(\phi(z)T^{-1}J_B(z)+\phi^{\dagger}J_A(z)\Big)\equiv
-\int dz \Big(\phi(z)J^{\dagger}(z) +h.c\Big),
\eeq
where
\beq
J(z)=T^{-1}J_A(z),\ \ J^{\dagger}(z)=T^{-1}J_B(z).
\eeq
The interaction part becomes rather complicated, involving several
operators $T$ or their inverse:
\beq
S_I=\frac{2\alpha_s^2N_c}{\pi}
\int dy\frac{d^2r_1d^2r_2d^2r_3}{r_{12}^2r_{23}^2r_{31}^2}
\Big( T^{-1}\phi^{\dagger}(y,\rho_{23})\cdot T^{-1}\phi{\dagger}(y,\rho_{13})
\cdot r_{12}^4T\phi(y,\rho_{12}) + h.c.\Big).
\eeq

With the physical meaning of operators $\phi$ and $\phi^{\dagger}$ well established and
indeed standard, the analysis of evolution becomes trivial.
Let us follow it  for free pomerons. Then their number is conserved and
actually the only connected diagram corresponds to a single pomeron.
Such a state is to be constructed as a superposition of single pomerons at
different positions $\rz$:
\beq
\Psi(y)=\int d\rz f(y,\rz)\phi^{\dagger}(\rz)\Psi_0,
\eeq
where $\Psi_0$ is the vacuum state which obeys
\beq
\phi(\rz)\Psi_0=0
\eeq
and is normalized to unity.
(we assume the Shroedinger-like picture with operators $\psi$ and $\psi^{\dagger}$
at fixed rapidity). At the initial rapidity $y=0$ the pomeron wave function
is determined by the external current:
\beq
\Psi(0)=\int d\rz \bar{J}^{\dagger}(\rho)\phi^{\dagger}(\rz)\Psi_0,
\eeq
where we recall that $\bar{J}$ is the spatial part of $J$.
This state evolves to the final rapidity $Y$ at which we are interested
in the amplitude $A_{fi}$ to pass to  a specific final state determined by the
current at $y=Y$:
\beq
\Psi_f=\int d\rz \bar{J}(\rz)\phi^{\dagger}(\rz)\Psi_0.
\eeq
One has
\beq
A_{fi}=<\Psi_f|\Psi(Y)>=\int d\rz d\rzp
\bar{J}(\rz)f(y,\rzp)<\Psi_0|\phi(\rz)
\phi^{\dagger}(\rzp)|\Psi_0>=
\int d\rz \bar{J}(\rz)f(y,\rz),
\eeq
where we used (35) and (44).

The law which governs evolution of the wave function $f(y,\rz)$ follows from the
general Schroedinger equation (31) and the form of the Hamiltonian ${\cal
H}$.
The free part of the latter in terms of new operators has the standard form
\beq
{\cal H}_0=
-\int  d\rho\phi^{\dagger}(\bar{z})\bar{H}\phi(\bar{z}),
\eeq
so that from (43) one immediately finds the equation
\beq
\frac{\partial f(y,\rz)}{\partial y}=-\bar{H}f(y,\rz),
\eeq
with a formal solution
\beq
f(z)=e^{-\bar{H} y}f(0)=\int d\rzp \bar{g}(y,\rz;0,\rzp)f(0).
\eeq
Here $\bar{g}$ is the Green function for the operator
$\partial/\partial y+\bar{H}$
which can be written as an operator in the coordinate space
\beq
\bar{g}(y)=\theta(y)e^{-\bar{H}y}.
\eeq
Using this we obtain for the amplitude
\beq
A_{fi}=\int d\rz d\rzp\bar{J}(\rz)\bar{g}(y,\rz;0,\rzp)\bar{J}^{\dagger}(\rzp).
\eeq
Returning  to the initial external sources and the Green function
we reproduce the standard result
\beq
A_{fi}=\int d\rz d\rzp \bar{J}_A(\rz)g(y,\rz;0,\rzp)\bar{J}_B(\rzp).
\eeq
Indeed we have
\beq
\bar{g}=\Big(\frac{\partial}{\partial y}+\bar{H}\Big)^{-1}=
T\Big(\frac{\partial}{\partial y}+H\Big)^{-1}T^{-1}=TgT
\eeq
Putting this into (52) gives (53).


\section{The Schwinger-Dyson equations for the pomeron Green function}
\subsection{The pomeron self-mass}

The pomeron self-mass operator starts and finishes with the
3-pomeron vertex, which contains operator $L$ acting on the
incoming and outgoing pomeron propagator. As a result, the Dyson
equation for the full Pomeron Green function $G(z,z')$ takes the form
\beq
G(z,z')=g(z|z')-\int d\tilde{z}_{inv}d\tilde{z}'_{inv}
g(z,\tilde{z})L
\Sigma(\tilde{z},\tilde{z}')
LG(\tilde{z}',z'),
\eeq
where both operators $L$ act on pomeron propagators
(the minus sign in front of the second term is due to the propagator
being in fact equal to $-g$).
Applying to this equation operators $L$ from the left and from the
right we find
\beq
\tilde{G}(z,z')=\tilde{g}(z,z')-\int d\tilde{z}_{inv}d\tilde{z}'_{inv}
\tilde{g}(z,\tilde{z})\Sigma(\tilde{z},\tilde{z}')
\tilde{G}(\tilde{z}',z'),
\eeq
where we define
\beq
\tilde{G}=LGL,\ \ \tilde{g}=LgL.
\eeq
Eq. (57) can be rewritten in an obvious operatorial form as
\beq
\tilde{G}=\tilde{g}-\tilde{g}\Sigma\tilde{G},
\eeq
which is the standard form for the Dyson equation, except for the sign.

Note that  the self-mass $\Sigma$ entering this equation is a conformal
invariant function, due to conformal invariance of both the pomeron
propagator and 3-pomeron vertex. As a result the full Green function
is also conformal invariant.

To pass to pomeron 'energies' $\omega$  we  have to understand how
they are related in the 3-pomeron vertex.
We standardly present
\beq
G(y)=\int_{a-i\infty}^{a+i\infty}\frac{d\omega}{2\pi i}e^{\omega y}G(\omega),
\eeq
with the inverse transform
\beq
G(\omega)=\int_{-\infty}^{+\infty} dy e^{-\omega y}G(y).
\eeq
In the lowest non-trivial order, suppressing the integration over coordinates
and coupling at the vertexes, we have
\beq
G(Y)=\int dy dy'dy''\delta(y+y'+y''-Y)g(y)g_1(y'')g_2(y'')g(y'),
\eeq
where $g_1$ and $g_2$ are the  propagators of the two intermediate
pomerons. For $G(\omega)$ we find
\beq
G(\omega)=\int dy dy'dy'' e^{-\omega(y+y'+y'')}g(y)g_1(y'')g_2(y'')g(y')=
g(\omega)\int dy e^{-\omega y}g_1(y)g_2(y)g(\omega).
\eeq
The integral over $y$ can be written in the form
\beq
\int dy e^{-\omega y}\int_{a_1-i\infty}^{a_1+i\infty}\frac{d\omega_1}{2\pi i}
\int_{a_2-i\infty}^{a_2+i\infty}\frac{d\omega_2}{2\pi i}
e^{y(\omega_1+\omega_2)}g_1(\omega_1)g_2(\omega_2).
\eeq
We can always choose $\omega$ on the line
\[ {\rm Re}\,(\omega-\omega_1-\omega_2)=0. \]
Then integration over $y$ will give
\[ 2\pi \delta({\rm Im}\,(\omega-\omega_1-\omega_2)).\]
Subsequent integration over, say, $\omega_2$ will finally give
\beq
\int dy e^{-\omega y}g_1(y)g_2(y)=
\int_{a_1-i\infty}^{a_1+i\infty}\frac{d\omega_1}{2\pi i}
g_1(\omega_1)g_2(\omega-\omega_2),
\eeq
which result should be analytically continued to arbitrary $\omega$.
This means that in terms of energies the 3-pomeron vertex
formally contains
 \beq
 2\pi i\delta(\omega-\omega_1-\omega_2).
 \eeq
that is, energies are conserved at the vertex.

Using this result, again in the lowest order, we find, explicitly
showing the coordinates: $(1,2)\equiv\{r_1,r_2\}$
\beq
\Sigma^{(0)}_\omega(1,2|1',2')=\frac{8\alpha_s^4N_c^2}{\pi^2}
\int\frac{d\omega_1}{2\pi i}\int \frac{r_{12}d^2r_3}{r_{31}^2r_{21}^2}
\frac{r'_{12}d^2r'_3}{{r'_{31}}^2{r'_{21}}^2} g_{\omega_1}(1,3|1',3')
g_{\omega-\omega_1}(2,3|2',3').
\eeq

To pass to the full self-mass we have to substitute the full pomeron Green
functions $G$ for the propagators $g$ and change one of the 3-Pomeron
vertices into the full one $\Gamma$. In this way we obtain
\[
\Sigma_\omega(1,2|1',2')=\frac{8\alpha_s^4N_c^2}{\pi^2}
\int\frac{d\omega_1}{2\pi i}\int \frac{r_{12}d^2r_3}{r_{31}^2r_{21}^2}
d\tau(1'',3'')d\tau(2'',4'')
\]\beq G_{\omega_1}(1,3|1'',3'')
G_{\omega-\omega_1}(2,3|2'',4'')\Gamma_{\omega,\omega_1}
(1'',2'',3'',4''|1',2').
\eeq
In the lowest order the vertex is given by the spatial part of $\gamma$,
Eq. (14):
\beq
\Gamma^{(0)}(1'',2'',3'',4''|1',2')=\delta^2(1''-1')\delta^2(2''-2')
\delta^2(3''-4'')r_{1'2'}^2r_{1''3''}^2r_{2''4''}^2.
\eeq

\subsection{The vertex equation}
As in any quantum field theory with a triple interaction the full
vertex is determined by an infinite sequence of skeleton diagrams,
containing the vertex itself and full Green functions. In the lowest order
one has the so-called  '3-gamma' equation. Its explicit form
is rather complicated due to large number of variables. From the
corresponding Feynman diagram we find in terms of gluon coordinates
\[
\Gamma_{\omega,\omega'}(1',2',3',4'|1,2)=
\Gamma^{(0)}(1',2',3',4'|1,2)\]\[
-\frac{8\alpha_s^4N_c^2}{\pi^2}
\int\frac{d\omega_1}{2\pi i}
\int d\tau(1'',2'')d\tau(3'',4'')d\tau(3''',4''')
d\tau(\tilde{1},\tilde{2})
d\tau(\tilde{1}',\tilde{2}')d\tau(1''',2''')\]
\[
L(1',3')\Gamma_{\omega',\omega_1}
(1',3'|1'',2'',3'',4'')G_{\omega_1}(1'',2''|1''',2''')
\]\[G_{\omega'-\omega_1}(3'',4''|3''',4''')
\Gamma_{\omega-\omega_1,\omega-\omega'}
(2',4',3''',4'''|\tilde{1},\tilde{2})\]\beq
L(\tilde{1},\tilde{2})G_{\omega-\omega_1}
(\tilde{1},\tilde{2}|\tilde{1}',\tilde{2}')
\Gamma_{\omega,\omega_1}(1''',2''',\tilde{1}',\tilde{2}'|1,2).
\eeq
In the shorthand notation $\rho=\{r_1,r_2\}$ and correspondingly
\[G_\omega(1,2)\equiv G_\omega(\rho_1,\rho_2)\equiv
G(r_1^{(1)},r_2^{(1)}r_1^{(2)}r_2^{(2)})\]
the equation
for the vertex part can be rewritten in a more compact form
\[
\Gamma_{\omega,\omega'}(1',2'|1)=\Gamma^{(0)}(1',2'|1)
-\frac{8\alpha_s^4N_c^2}{\pi^2}
\int\frac{d\omega_1}{2\pi i}
\int d\tau(1'')d\tau(2'')d\tau(3'')
d\tau(4'')
d\tau(5'')d\tau(6'')\]\[L(1')
\Gamma_{\omega',\omega_1}(1'|1'',2'')
G_{\omega_1}(1''|6'')G_{\omega'-\omega_1}(2''|3'')\]\beq
\Gamma_{\omega-\omega_1,\omega-\omega'}(2',3''|4'')
L(4'')G_{\omega-\omega_1}(4''|5'')
\Gamma_{\omega,\omega_1}(5'',6''|1).
\eeq

\section{Conformal (Moebius) invariance}
\subsection{Conformal basis}
The basic building blocks of the pomeron perturbation theory:
the propagator $g$, bare 3-pomeron vertex $\gamma$ and
integration volume $d\tau$, - are conformal invariant.
As a result also the full Green function $G$, self-mass $\Sigma$
and vertex $\Gamma$ are conformal invariant functions of their
arguments.  It seems profitable to use this property to simplify
the Schwinger-Dyson equations. To do this we have to study a
general form for the conformal invariant functions for the
transitions pomeron$\to$ pomeron ('two-point functions')
and pomeron$\to$ 2 pomerons ('three-point functions').
This can be achieved by expanding these functions in the
conformal basis formed by functions
(in complex notation) ~\cite{lip}
\beq
E_\mu(\rho)=E_{\mu}(r_1,r_2)
=\left(\frac{r_{12}}{r_{10}r_{20}}\right)^{\frac{1-n}{2}+i\nu}
\left(\frac{r^*_{12}}{r^*_{10}r^*_{20}}\right)^{\frac{1+n}{2}+i\nu},
\eeq
where $\mu=\{n,\nu,r_0\}=\{h,r_0\}$ with $n$ integer, $\nu$ real and
two-dimensional transverse $r_0$ enumerate the basis.

Functions $E_{\mu}(\rho)$ are the proper functions of the operator
$L$
\beq
L\,E_{\mu}(\rho)=l_\mu E_{\mu}(\rho), \ \ l_\mu=\frac{4\pi^8}
{a_{n+1,\nu}a_{n-1,\nu}},
\eeq
where
\beq
a_{n,\nu}\equiv a_{\mu}=\frac{\pi^4}{2}\frac{1}{\nu^2+n^2/4}.
\eeq
They form a complete system:
\beq
r_{12}^4\delta(\rho-\rho')=
\frac{1}{2}\sum_{\mu}
E_{\mu}(\rho)E^*_{\mu}(\rho'),
\eeq
where we use a notation
\beq
\sum_{\mu}=\sum_{n=-\infty}^{\infty}\int d\nu\frac{1}{a_{n,\nu}}
 \int d^2r_0,
\eeq
and satisfy the orthogonality relation
\[
\int d\tau
E_{\mu}(\rho)E^*_{\mu'}(\rho')\]\beq
=a_{n,\nu}\delta_{nn'}\delta(\nu-\nu')\delta^2(r_{00'})+
b_{n\nu}\delta_{n,-n'}\delta(\nu+\nu')|r_{00'}|^{-2-4i\nu}(\frac{r_{00'}}
{r^*_{00'}})^n.
\eeq
The coefficients $b_{n\nu}$ are given by the formula
\beq
b_{n\nu}=\pi^3\frac{2^{4i\nu}}{-i\nu+|n|/2}
\frac{\Gamma(-i\nu+(1+|n|)/2)\Gamma(i\nu+|n|/2)}
{\Gamma(i\nu+(1+|n|)/2)\Gamma(-i\nu+|n|/2)}.
\eeq
Functions $E_{\mu}$ are not linearly independent. In fact
\beq
E_{-n,-\nu,r_0}(\rho)=\frac{b_{n\nu}}{a_{n\nu}}\int d^2r'_0
|r_{00'}|^{-2+4i\nu}\left(\frac{r_{00'}^*}{r_{00'}}\right)^n E_{n,\nu,r'_0}
(\rho).
\eeq

Presenting a function $f(\rho)$  as
\beq
f(\rho)=\int d\rho'\delta(\rho-\rho')f(\rho')
\eeq
and using completeness (74) one gets
\beq
f(\rho)=\frac{1}{2}\int d\tau'
\sum_{\mu}E_{\mu}(\rho)E^*_{\mu}(\rho')f(\rho')
=\frac{1}{2}\sum_{\mu}E_{\mu}(\rho)f_{\mu},
\eeq
where
\beq
f_{\mu}=\int d\tau E^*_{\mu}(\rho)f(\rho).
\eeq
This gives the standard expansion in the whole overcomplete basis.

However it seems possible to limit oneself to an independent part of this
basis. One possibility is to take a restricted basis
with $\nu>0 (\nu<0)$. We denote this restriction by $\mu>0 (\mu<0)$.
In fact we may split the integration over $\nu$ in (80)  into two parts:
\beq
f(\rho)
=\frac{1}{2}\sum_{\mu>0}E_{\mu}(\rho)f_{\mu}+
\frac{1}{2}\sum_{\mu<0}E_{\mu}(\rho)f_{\mu}
\eeq
and, say, in the second term express the eigenfunctions $E_{\mu}$
with $\mu<0$ via those with $\mu>0$ using (78)
to obtain
for the second term
\beq
\frac{1}{2}\sum_{\mu<0}f_{\mu}
\frac{b_{-n-\nu}}{a_{-n-\nu}}\int d^2r'_0
|r_{00'}|^{-2-4i\nu}\left(\frac{r_{00'}^*}
{r_{00'}}\right)^{-n} E_{-n,-\nu,r'_0}(\rho).
\eeq
Changing summation and integration variables $n\to -n$ and $\nu\to -\nu$
we get for this term
\beq
\frac{1}{2}\sum_{\mu>0}f_{-n,-\nu,r_0}
\frac{b_{n\nu}}{a_{n\nu}}\int d^2r'_0
|r_{00'}|^{-2+4i\nu}\left(\frac{r_{00'}^*}
{r_{00'}}\right)^{n} E_{n,\nu,r'_0}(\rho).
\eeq
Interchanging also $r_0$ and $r'_0$ integrations we finally find for it
\beq
\frac{1}{2}\sum_{\mu>0}E_{\mu}(\rho)\bar{f}_{\mu},
\eeq
where
\beq
\bar{f}_{\mu}=\frac{b_{n\nu}}{a_{n\nu}}
\int d^2r'_0f_{-n,-\nu,r'_0}
|r_{0'0}|^{-2+4i\nu}\left(\frac{r_{0'0}^*}
{r_{0'0}}\right)^{n}.
\eeq
Summing this with the first term in (82) we get the desired expansion
in states with $\mu>0$
\beq
f(\rho)=\frac{1}{2}\sum_{\mu>0}E_{\mu}(\rho)
(f_{\mu}+\bar{f}_{\mu})\equiv
\sum_{\mu>0}E_{\mu}(\rho)\lambda_{\mu}.
\eeq
Integrating this with $E^*_{\mu>0}(\rho)$ and using (76) we find
\beq
\lambda_{\mu}=
\int d\tau
E^*_{\mu}(\rho)f(\rho).
\eeq
Putting this into (87) we find
\beq
f(\rho)=
\sum_{\mu>0}E_{\mu}(\rho)
\int d\tau'
E^*_{\mu}(\rho')f(\rho'),
\eeq
which means that one also has a completeness relation for half of the
basis with $\nu>0$:
\beq
r_{12}^4\delta(\rho-\rho')=
\sum_{\mu>0}E_{\mu}(\rho)E^*_{\mu}(\rho').
\eeq
Obviously the same property is valid for the second half of
the basis with
$\nu<0$.

\subsection{Two-point functions}
We have to deal with a conformal invariant
function $A(1|1')$, where, as introduced in the previous sections,
the arguments refer to pairs of the pomeron coordinates.
Using half of the conformal basis we present $A$ as
\beq
A(1|1')=\sum_{\mu,\mu'>0}E_{\mu}(1)E^*_{\mu'}(1')A_{\mu\mu'}
=\sum_{\mu>0}\sum_{\mu'<0}E_{\mu}(1)E_{\mu'}(1')A_{\mu\mu'},
\eeq
where we use
\beq
E^*_{n,\nu,r_0}(\rho)=E_{-n,-\nu,r_0}(\rho).
\eeq
Our aim is to see which  properties  $A_{\mu,\mu'}$ should have
for the function $A(1,1')$ to be conformal invariant.
Since $E_{n,\nu,r_0}(r_1,r_2)=E_{n,\nu,r_0+a}(r_1+a,r_2+a)$
the translational invariance requires $A_{\mu\mu'}$ to depend only
on the difference $r_0-r'_0\equiv r_{00'}$.
Under inversion the expression to be summed over $\mu>0$ and
$\mu'<0$ changes as follows
\beq
A_{\mu\mu'}(r_{00'})\to
r_0^{-1-n+2i\nu}{r'_0}^{-1-n'+2i\nu'}\cdot(a.\ f.)\cdot
A_{\mu\mu'}\left(\frac{r_{00'}}{r_0r'_0}\right).
\eeq
Here and in the following (a. f.) means 'antiholomorhic factor', that is the
complex conjugate of the preceding factor.
Invariance under inversion requires that
\beq
A_{\mu\mu'}\left(\frac{r_{00'}}{r_0r'_0}\right)=
r_0^{1+n-2i\nu}{r'_0}^{1+n'-2i\nu'}\cdot (a.\ f.)\cdot
A_{\mu\mu'}(r_{00'}).
\eeq
However the left-hand side only depends on the product $r_0r'_0$,
so in the right-hand side we are obliged  to have either $n=n'$ and
$\nu=\nu'$ or  $A_{\mu\mu'}(r_{00'})\propto \delta(r_{00'})$.
Since $\nu$ and $\nu'$ have opposite signs, the first alternative
cannot be realized. With $A_{\mu\mu'}(r_{00'})\propto \delta(r_{00'})$.
we find that $n+n'=0$ and $\nu+\nu'=0$ so that
\beq
A_{\mu\mu'}=\delta_{n,-n'}\delta(\nu+\nu')\delta(r_{00'})a_{n\nu}A_{n\nu}
\equiv \delta_{\mu,\bar{\mu}'}A_{\mu},
\eeq
where $\bar{\mu}=\mu(n\to -n,\nu\to -\nu)$ and
we defined $A_\mu\equiv A_{n\nu}$ with factor $a_{n\nu}$ separated for convenience.
As a result the double sum in (91) transforms into a single one
\beq
A(1|1')=\sum_{\mu>0}E_{\mu}(1)E_{\bar{\mu}}(1')A_{n\nu}
=\sum_{\mu>0}E_{\mu}(1)E^*_{\mu}(1')A_{n\nu}
\eeq
where $\bar{\mu}=\mu(n\to -n,\nu\to -\nu)$.
A similar form with summation over $\mu<0$ can be obtained in the same
manner.
Representation (96) or a similar one with a sum over $\mu<0$ are valid for any
conformal invariant two-point function.
Note that taking an average of the sums over $\mu>0$ and $\mu<0$ one obtains
a similar representation in terms of the whole overcomplete basis, which is
standardly used for the BFKL Green function $g(z,z')$

Now
suppose we have conformal invariant functions
$B(1|1')$ and $C(1,1')$ and form a conformal invariant integral
\beq
A(1|1')=\int d\tau''B(1|1'')C(1''|1').
\eeq
Each of the three functions $A,B$ and $C$ has the representation (96)
with conformal coefficients $A_{\mu}$, $B_{\mu}$ and $C_{\mu}$.
Doing the integration with the
help of orthonormalization properties of the basis functions $E_{\mu}$
with $\mu>0$ we
find
\beq
A(1|1')=\sum_{\mu>0}E_{\mu}(1)E^*_{\mu}(1')B_{\mu}C_{\mu}
=\sum_{\mu>0}E_{\mu}(1)E^*_{\mu}(1')A_\mu,
\eeq
which means that in the conformal representation (96)
\beq
A_\mu=B_\mu C_\mu.
\eeq

Applying this result to the Dyson equation (58) in the $\omega$
representation we immediately find
\beq
\tilde{G}_{\omega\mu}=\tilde{g}_{\omega\mu}-
\tilde{g}_{\omega\mu}\Sigma_{\omega\mu}\tilde{G}_{\omega\mu}.
\eeq
So both the energy $\omega$ and conformal quantum numbers $\mu$ of the
pomeron are conserved in the interaction and  its full Green function in
the conformal basis is trivially expressed via its self-mass:
\beq
\tilde{G}_{\omega\mu}=\frac{1}{1/\tilde{g}_{\omega\mu}+
\Sigma_{\omega\mu}}, \ \ {\rm or}\ \
G_{\omega\mu}=\frac{1}{1/g_{\omega\mu}+
l_\mu^2\Sigma_{\omega\mu}}.
\eeq
In Eq.(101) the free pomeron Green function $g$
in the conformal basis is given by
\beq
g_{\omega\mu}=\frac{2}{l_{n\nu}}\,\frac{1}{\omega-\omega_{n\nu}}.
\eeq

\subsection{3-point functions}
For a 3-point function the expansion similar to (96) reads
\beq
\Gamma(1|2,3)=\sum_{\mu_1,\mu_2\mu_3>0}E_{\mu_1}(1)E^*_{\mu_2}(2)
E^*_{\mu_3}(3)\Gamma_{\mu_1|\mu_2\mu_3},
\eeq
where the intermediate c.m. coordinates are $R_1$, $R_2$
and $R_3$.
Conformal invariance allows to determine the dependence on them
of $\Gamma_{\mu_1|\mu_2\mu_3}$.
Translational invariance requires  $\Gamma_{\mu_1|\mu_2\mu_3}$ to
depend only on differences $R_{ik}$ and from the scale invariance
such a dependence should be a power one.
So we seek
\beq
\Gamma_{\mu_1|\mu_2\mu_3}=R_{12}^{\alpha_{12}}R_{23}^{\alpha_{23}}
R_{31}^{\alpha_{31}}\cdot
(a.\ f.)\cdot
\Gamma_{n_1\nu_1|n_2\nu_2n_3\nu_3}.
\eeq
After inversion we find an extra factor
in the sum (102)
\beq
R_1^{-2-\alpha_{12}-\alpha_{31}+1-n_1+2i\nu_1}\cdot
(a.\ f.)\cdot e^{-in_1 \pi}
\eeq
times two similar factors which are obtained from (104) by cyclic permutations
of 123 and conjugation. Invariance requires that each power is zero
and that the sum $n_1+n_2+n_3$ be even.
We get six equations to determine the $\alpha$'s. Their solution gives
\[
\alpha_{12}=-\frac{1}{2}+\frac{1}{2}(n_2-n_1-n_3)+i(\nu_1-\nu_2+\nu_3),\]\[
\alpha_{23}=-\frac{1}{2}+\frac{1}{2}(n_1+n_2+n_3)-i(\nu_1+\nu_2+\nu_3),\]
\beq
\alpha_{31}=-\frac{1}{2}+\frac{1}{2}(n_3-n_1-n_2)+i(\nu_1+\nu_2-\nu_3)
\eeq
and similar expressions for the powers in the
antiholomorhic factor $\tilde{\alpha}$'s with $n_i\to -n_i$.

For the free vertex $\Gamma^{(0)}$ given by Eq. (68) we have
\beq
\Gamma^{(0)}_{\mu_1|\mu_2\mu_3}=V_{\bar{\mu}_1\mu_2\mu_3},
\eeq
where again
 $\bar{\mu}=\mu(n\to-n,\nu\to-\nu)$ and
the vertex $V_{\mu_1\mu_2\mu_3}$ has been  introduced by Korchemsky
\cite{kor}:
\beq
V_{\mu_1\mu_2\mu_3}=\int\frac{d^2r_1d^2r_2d^2r_3}{r_{12}^2r_{23}^2r_{31}^2}
E_{\mu_1}(r_1,r_2)E_{\mu_2}(r_2,r_3)E_{\mu_3}(r_3,r_1)=
\Omega(h_1,h_2,h_3)
\prod_{i<j}r_{0_i0_j}^{-\Delta_{ij}}{r^*_{0_i0_j}}^{-\bar{\Delta}_{ij}}
\eeq
with $\Delta_{12}=h_1+h_2-h_3$ etc.

\section{$\Sigma$ and $\Gamma$ in the conformal basis}
To illustrate the simplifications introduced by  transition to the conformal
basis
in this section we study the pomeron self-mass and triple interaction vertex
in this basis

In a shorthand notation for the gluon coordinates $1=(r_1,\bar{r}_1)$
the pomeron self mass is given by
\[
\Sigma_\omega(1|1')=\frac{8\alpha_s^4N_c^2}{\pi^2}
\int\frac{d\omega_1}{2\pi i}\int
d\tau(2)d\tau(3)d\tau(2')d\tau(3')\]\beq
\Gamma^{(0)}(1|2,3)G_{\omega_1}(2|2')G_{\omega-\omega_1}(3|3')
\Gamma_{\omega,\omega_1}(2',3'|1').
\eeq
The actual number of integrations is in fact  smaller due to
$\delta$-functions in the conformal vertex $\Gamma^{(0)}$ defined by Eq. (68).
We expand both the Green function and the vertexes in the conformal basis
with $\mu>0$ (see 96) and (103). In the following we also
suppress  the $\omega$-dependence to economize on subindexes.
We have
\beq
G(2|2')=\sum_{\mu >0}G_{\mu}E_{\mu}(2)E^*_{\mu}(2'),
\eeq
\beq
G(3|3')=\sum_{\mu'>0}G_{\mu'}E_{\mu'}(3)E^*_{\mu'}(3'),
\eeq
\beq
\Gamma^{(0)}(1|2,3)=\sum_{\mu_{1},\mu_{2},\mu_{3}>0}
\Gamma^{(0)}_{\mu_{1}|\mu_{2},\mu_{3}}
E_{\mu_{1}}(1)E^*_{\mu_{2}}(2)E^*_{\mu_{3}}(3),
\eeq
\beq
\Gamma(2',3'|1')=\sum_{\mu'_{1},\mu'_{2},\mu'_{3}>0}
\Gamma_{\mu'_{2},\mu'_{3}|\mu'_{1}}
E^*_{\mu'_{1}}(1')E_{\mu'_{2}}(2')E_{\mu'_{3}}(3').
\eeq
Integrations over the gluon coordinates are done with the help of (76) and give
the product of $\delta$-functions:
\[
a_\mu a_{\mu'}\delta_{\mu\mu_{2}}\delta_{\mu\mu'_{2'}}
\delta_{\mu'\mu_{3}}\delta_{\mu'\mu_{3'}}.\]
So we find
\[
\Sigma (1,2|1',2')=\frac{8\alpha_s^4N_c^2}{\pi^2}
\int\frac{d\omega_1}{2\pi i}
\sum_{\mu_{1},\mu'_{1},\mu,\mu'}\Gamma^{(0)}_{\mu_{1}|\mu,\mu'}
G_{\mu}G_{\mu'}
\Gamma_{\mu,\mu|\mu'_{1}}E_{\mu_{1}}(1)E^*_{\mu'_{1}}(1')
\]\beq=
\sum_{\mu_{1},\mu'_{1}}\Sigma_{\mu_{1}\mu'_{1}}
E_{\mu_{1}}(1)E^*_{\mu'_{1}}(1'),
\eeq
where
\beq
\Sigma_{\mu_{1}\mu'_{1}}=
\frac{8\alpha_s^4N_c^2}{\pi^2}
\int\frac{d\omega_1}{2\pi i}
\sum_{\mu,\mu'}\Gamma^{(0)}_{\mu_{1}|\mu,\mu'}
G_{\mu}G_{\mu'}
\Gamma_{\mu,\mu|\mu'_{1}}.
\eeq
and the suppressed  dependence on $\omega$ is obvious from its conservation
at the vertexes.
Thus we have found for $\Sigma(1,2|1',2')$ an expansion in the conformal
basis (91). From its conformal invariance it follows that
\beq
\Sigma_{\mu_{1}\mu'_{1}}=\delta_{\mu_{1}\mu'_{1}}
\Sigma_{\mu_{1}},
\eeq
where $\Sigma_{\mu}$ is the desired pomeron self mass in the
conformal representation. It can be found from (115) after summation over $\mu'_{1}$
This gives
\beq
\Sigma_{\mu}=
\frac{8\alpha_s^4N_c^2}{\pi^2}
\int\frac{d\omega_1}{2\pi i}
\sum_{\mu_{1},\mu_2,\mu_3}\Gamma^{(0)}_{\mu|\mu_1,\mu_2}
G_{\mu_1}G_{\mu_2}
\Gamma_{\mu_1,\mu_2|\mu_3}.
\eeq

The sum over $\mu_i$, $i=1,2,3$ includes integrations over three c.m coordinates
$R_i$ on which only the vertexes depend. We get an integral
depending on the four conformal weights
\beq
I_{h|h_1,h_2|h_3}=
\int d^2R_1d^2R_2d^2R_3R_{01}^{\alpha_{01}}R_{02}^{\alpha_{02}}R_{12}^{\alpha_{12}}
R_{31}^{\alpha_{31}}R_{32}^{\alpha_{32}}R_{12}^{\alpha_{21}}\cdot \Big( a.f.\Big).
\eeq
Here $R_0$ is arbitrary since the integral is independent of it.
The powers are given by
\[\alpha_{01}=-\frac{1}{2}+\frac{1}{2}(n_1-n_2-n)+i(\nu_2-\nu_1-\nu),\]
\[\alpha_{12}=-\frac{1}{2}+\frac{1}{2}(n_1+n_2+n)-i(\nu_2+\nu_1+\nu),\]
\[\alpha_{21}=-\frac{1}{2}-\frac{1}{2}(n_1+n_2+n_3)+i(\nu_2+\nu_1+\nu_3),\]
\[\alpha_{31}=-\frac{1}{2}+\frac{1}{2}(n_2-n_1+n_3)+i(\nu_1-\nu_2-\nu_3)\]
and $\alpha_{02}$ and $\alpha_{32}$ are obtained from
$\alpha_{01}$ and $\alpha_{31}$ by  interchange 1$\leftrightarrow $2.
The integral (118) is convergent both in the infrared and ultraviolet.
However its calculation does not look simple.
Once this integral is known, the self-mass in the conformal basis
is given by a sum over three conformal weights:
 \beq
\Sigma_{\omega,h}(\omega)=
\frac{8\alpha_s^4N_c^2}{\pi^2}
\int\frac{d\omega_1}{2\pi i}
\sum_{h_{1},h_2,h_3}I_{h|h_1,h_2|h_3}\Omega_{\bar{h},h_1,h_2}
G_{h_1}(\omega_1)G_{h_2}(\omega-\omega_1)
\Gamma_{\bar{h}_1,\bar{h}_2|h_3}(\omega,\omega_1).
\eeq
Here we made explicit the $\omega$ dependence introducing it in the
arguments;
$\Omega$ is the Korchemski vertex (107), $\Gamma_{\bar{h}_1,\bar{h}_2|h_3}$
is defined by Eq. (104) and
$\sum_h$ is given by (74) without integration over $r_0$.

In the same manner one can obtain expressions for the vertex part $\Gamma$.
We shall limit ourselves with the '3-gamma' approximation, Eq. (79).
As before we expand the vertex parts and Green functions in the conformal
basis. Suppressing again the $\omega$-dependence we have
\[
\Gamma(1'|1'',2'')=
\sum_{\mu_3,\mu'_3,\mu''_3>0}\Gamma_{\mu_3|\mu'_3,\mu''_3}
E_{\mu_3}(1')E^*_{\mu'_3}(1'')E^*_{\mu''_3}(2''),
\]
\[
\Gamma(2',3''|4'')=\sum_{\mu_2,\mu'_2,\mu''_2>0}\Gamma_{\mu_2\mu'_2|\mu''_2}
E^*_{\mu''_2}(4'')E_{\mu_2}(2')E_{\mu'_2}(3''),
\]
\[
\Gamma(5'',6''|1)=\sum_{\mu_1,\mu'_1,\mu''_1>0}\Gamma_{\mu'_1\mu''_1|\mu_1}
E^*_{\mu_1}(1)E_{\mu'_1}(5'')E_{\mu''_1}(6''),
\]
\[G(1''|6'')=\sum_{\mu_4>0}G_{\mu_4}E_{\mu_4}(1'')E^*_{\mu_4}(6''),\]
\[G(2''|3'')=\sum_{\mu_5>0}G_{\mu_5}E_{\mu_5}(2'')E^*_{\mu_5}(3''),\]
\[G(4''|5'')=\sum_{\mu_6>0}G_{\mu_6}E_{\mu_6}(4'')E^*_{\mu_6}(5'').\]
Integrations over the double primed coordinates will give 6 $\delta$-functions
in $\mu$. So finally we are left with 6 summations over $\mu_i$, i=1,...6.
The result can be presented in the form (with $\omega$-dependence suppressed)
\beq
\Gamma(1',2'|1)=
\sum_{\mu_1,\mu_2,\mu_3>0}\Gamma_{\mu_2,\mu_3|\mu_1}
E^*_{\mu_1}(1)E_{\mu_2}(1')E_{\mu_3}(2'),
\eeq
where
\beq
\Gamma_{\mu_2,\mu_3|\mu_1}=\Gamma^{(0)}_{\mu_2,\mu_3|\mu_1}
-\frac{8\alpha_s^4N_c^2}{\pi^2}
\int\frac{d\omega_1}{2\pi i}
l_{\mu_3}\sum_{\mu_4,\mu_5,\mu_6>0}l_{\mu_6}\Gamma_{\mu_3|\mu_4,\mu_5}
\Gamma_{\mu_2,\mu_5|\mu_6}\Gamma_{\mu_6,\mu_4|\mu_1}
G_{\mu_4}G_{\mu_5}G_{\mu_6}
\eeq
is the desired vertex part in the conformal basis. 
Its dependence
on the c.m. coordinates $R_i$, $i=1,2,3$, is determined according to
Eq. (104). The bare vertex
$\Gamma^{(0)}_{\mu_2,\mu_3|\mu_1}$ is here given by the formula 
analogous to (107).

In (121) summations over $\mu_i$, $i=4,5,6$, include integrations over c.m.
coordinates $R_i$, $i=4,5,6$. In the integrand the $R$ dependence comes only from
the vertex parts and in its turn is defined by Eq. (104).
Thus the expression
\[
I_{h_1h_2h_3|h_4h_5h_6}\]\beq=
R_{12}^{-\alpha_{12}}R_{13}^{-\alpha_{13}} R_{23}^{-\alpha_{23}}
\int dR_4dR_5dR_6
R_{34}^{\alpha_{34}} R_{35}^{\alpha_{35}} R_{45}^{\alpha_{45}}
R_{14}^{\alpha_{14}} R_{16}^{\alpha_{16}} R_{46}^{\alpha_{46}}
R_{25}^{\alpha_{25}} R_{26}^{\alpha_{26}} R_{56}^{\alpha_{56}}
\cdot \Big( a.f. \Big),
\eeq
where all $\alpha$'s are determined by formulas similar to (106)
(see Appendix 2.),
is independent of c.m. coordinates  $R_i$, $i=1,2,3$ and depends only
on conformal weights. So the part of the vertex depending on conformal weights
will satisfy an equation (with $\omega$ dependence restored in arguments)
\[
\Gamma_{h_2,h_3|h_1}(\omega,\omega')=\Omega(h_1,\bar{h}_2,\bar{h}_3)
-\frac{8\alpha_s^4N_c^2}{\pi^2}
\int\frac{d\omega_1}{2\pi i}
l_{h_3}\sum_{h_4,h_5,h_6>0}l_{h_6}
I_{h_1h_2h_3|h_4h_5h_6}\]\beq
\Gamma_{h_3|h_4,h_5}(\omega',\omega_1)
\Gamma_{h_2,h_5|h_6}(\omega-\omega_1,\omega-\omega')
\Gamma_{h_6,h_4|h_1}(\omega,\omega_1)
G_{h_4}(\omega_1)G_{h_5}(\omega'-\omega_1)G_{h_6}(\omega-\omega_1).
\eeq

\section{Conclusions}
We have presented a formalism which allows to study interaction of pomerons
in the QCD with $N_c\to\infty$ using the standard methods of quantum field
theory. In particular we constructed the Schwinger-Dyson equations which
sum the so-called enhanced graphs and carry information of the 'physical'
pomeron as compared to the 'bare' one.

Conformal symmetry of the theory
leads to certain simplifications. As a result we obtain a picture very
similar to old Gribov local supercritical pomeron. The difference  is reduced
to an (infinite) number of pomerons with varying $n=0,\pm 2, \pm 4,...$ and
 more complicated form of the 'energy' $\omega_{n\nu}$ as a function
of $\nu$, which plays the role of the pomeron momentum in the old theory,
and of the  bare triple pomeron vertex, which now depends both on
$n$ and $\nu$. If however one selects the supercritical pomeron with $n=0$
and small values of $\nu$ the formal similarity becomes almost complete,
since the energy becomes a quadratic function of $\nu$ and
the bare vertex then reduces to a well-know constant ~\cite{kor}.
Unfortunately with this similarity also the problems of the old theory,
mentioned in the Introduction, return together with the question of the
internal consistency of the model. At present we do not know the answer to
this question and leave it for future studies.

\section{Acknowledgments}
The author has benefited from fruitful discussions with J.Bartels.
He is indebted for hospitality to  Hamburg University where this
paper was finished. This work has been supported by the NATO grant
PST.CLG.980287.

\section{Appendix 1. Color factors}
Let the color wave function of a pair of gluons be $|ab\rangle$, where
$a,b= 1,... N_c^2-1$. Then the vacuum color state is obviously
\beq
|0\rangle=\frac{1}{\sqrt{N_c^2-1}}\sum_a|aa\rangle=P|ab\rangle
\eeq
where
\beq
P=\frac{1}{\sqrt{N_c^2-1}}\delta_{ab}
\eeq
is the projector onto the vacuum color state.

The color structure of the vertex for the transition 2$\to $4  reggeized
gluons, with initial and final color variables $a_1,b_1\to a_2,b_2,a_3,b_3$
is given by the expression
\beq
V_c=f^{a_1a_2c}f^{cb_2d}f^{da_3e}f^{eb_3b_1}
\eeq
We want the projection of this color vertex onto the 3 vacuum color states
formed by the gluons with colors $a_1b_1$, $a_2b_2$ and $a_3b_3$.
Applying three corresponding projectors $P_1$, $P_2$ and $P_3$ we obtain
\beq
P_2P_3V_cP_1=\frac{1}{(N_c^2-1)^{3/2}}
f^{a_1a_2c}f^{ca_2d}f^{da_3e}f^{ea_3a_1}=
\frac{N_c^2}{(N_c^2-1)^{3/2}}\delta_{a_1d}\delta_{a_1d}=\frac{N_c^2}
{\sqrt{N_c^2-1}}\simeq N_c
\eeq
Note however that
the quarks quark loop which represents the external source
has its
color factor $\delta_{ab}=\sqrt{N_c^2-1} P\simeq N_cP$, so that
each external source contributes a factor $N_c$.

\section{Appendix 2. Powers in  Eq. (122)}
\[\alpha_{12}=-\frac{1}{2}+\frac{1}{2}(n_1+n_3-n_2)+i(\nu_2-\nu_3-\nu_1)\]
\[\alpha_{13}=-\frac{1}{2}+\frac{1}{2}(n_1+n_2-n_3)+i(\nu_3-\nu_2-\nu_1)\]
\[\alpha_{23}=-\frac{1}{2}-\frac{1}{2}(n_1+n_2+n_3)+i(\nu_1+\nu_2+\nu_3)\]
\[\alpha_{34}=-\frac{1}{2}+\frac{1}{2}(n_4-n_5-n_3)+i(\nu_5-\nu_4+\nu_3)\]
\[\alpha_{35}=-\frac{1}{2}+\frac{1}{2}(n_5-n_4-n_3)+i(\nu_4-\nu_5+\nu_3)\]
\[\alpha_{45}=-\frac{1}{2}+\frac{1}{2}(n_4+n_5+n_3)-i(\nu_5+\nu_4+\nu_3)\]
\[\alpha_{16}=-\frac{1}{2}+\frac{1}{2}(n_1+n_4-n_6)+i(\nu_6-\nu_4+\nu_1)\]
\[\alpha_{14}=-\frac{1}{2}+\frac{1}{2}(n_1+n_6-n_4)+i(\nu_4-\nu_6+\nu_1)\]
\[\alpha_{46}=-\frac{1}{2}-\frac{1}{2}(n_1+n_4+n_6)+i(\nu_6+\nu_4+\nu_1)\]
\[\alpha_{26}=-\frac{1}{2}+\frac{1}{2}(n_2+n_5-n_6)+i(\nu_6-\nu_5-\nu_2)\]
\[\alpha_{25}=-\frac{1}{2}+\frac{1}{2}(n_2+n_6-n_5)+i(\nu_5-\nu_6-\nu_2)\]
\[\alpha_{56}=-\frac{1}{2}-\frac{1}{2}(n_2+n_5+n_6)+i(\nu_6+\nu_5+\nu_2)\]


\begin{thebibliography}{99}
%
\bibitem{bal} I.Balitsky,  Nucl. Phys. {\bf B 463} (1996) 99.
%
\bibitem{kov} Yu.Kovchegov, Phys. Rev. {\bf D 60} (1999) 034008;
{\bf D 61} (2000) 074018.
%
\bibitem{bra1} M.A.Braun, Eur. Phys. J {\bf C 16} (2000) 337.
%
\bibitem{bra2} M.A.Braun, Phys. Lett. {\bf B 483} (2000) 115;
Eur. Phys. J {\bf C 33} (2004) 113; hep-ph/0504002 (to be published
in Phys.Lett. B)
%
\bibitem{salam} G.Salam, Nucl. Phys. {\bf B 461} (1996) 512
%
\bibitem{BRV} J.Bartels, M.Ryskin and G.P. Vacca, Eur. Phys. J. {\bf C 27}
(2003) 101.
%
\bibitem{jimwlk} J.-P. Blaizot, E.Iancu, K.Itakura and D.N.Triantafyllopulos,
Phys. Lett. {\bf B 615} (2005) 221.
%
\bibitem{mul} A.H.Mueller, A.I.Shoshi and S.M.H.Wong, Nucl. Phys.
{\bf B 715} (2005) 440.
%
\bibitem{abarb} H.D.I.Abarbanel, J.B.Bronzan, A.Schwimmer and R.L.Sugar,
Phys. Rev. {\bf D 14} (1976) 632.
%
\bibitem{white} A.White, Nucl. Phys. {\bf B 159} (1979) 77.
%
\bibitem{lip0} L.N.Lipatov, Sov.Phys. JETP {\bf 63} (1986) 904;
Nucl. Phys. {\bf B 365} (1991) 641.
%
\bibitem{BW} J.Bartels, Z.Phys. {\bf C 60} (1993) 471; 
J.Bartels and M.Wuesthoff, Z.Phys. {\bf C 66} (1995) 157.
%
\bibitem{AHM} A.H.Mueller and B.Patel, Nucl. Phys. {\bf B 425} (1994) 471.
%
\bibitem{BV} M.A.Braun and G.P.Vacca, Eur. Phys. J {\bf C6} (1999) 147.
%
\bibitem{BLV} J.Bartels, L.N.Lipatov and G.P.Vacca, Nucl. Phys.
{\bf B 706} (2005) 391.
%
\bibitem{lip} L.N.Lipatov, in "Perturbative QCD", ed. A.H.Mueller,
World. Sci. Singapore (1989).
%
\bibitem{kor} G.Korchemsky, Nucl. Phys. {\bf B 550} (1999) 397.
%
\end{thebibliography}
\end{document}